\documentclass[twocolumn,pra]{revtex4}
\usepackage{graphicx}
\usepackage{dcolumn}
\usepackage{bm}
\begin{document}

\preprint{APS/123-QED}
\title{Spin squeezing and entanglement in spinor-1 condensates}
\author{\"Ozg\"ur E. M\"ustecapl{\i}o\u{g}lu, M. Zhang, and L. You}
\address{School of Physics, Georgia Institute of Technology, Atlanta GA
30332, USA}
\date{\today}

\begin{abstract}
We analyze quantum correlation properties of a spinor-1
($f=1$) Bose Einstein condensate using the Gell-Mann realization
of SU(3) symmetry. We show that previously discussed phenomena
of condensate fragmentation and spin-mixing can be explained in
terms of the hypercharge symmetry. The ground state of a spinor-1
condensate is found to be fragmented for ferromagnetic interactions.
The notion of two bosonic mode squeezing
is generalized to the two spin ($U$-$V$) squeezing within the SU(3) formalism.
Spin squeezing in the isospin subspace ($T$) is found and numerically
investigated. We also provide new results for the stationary
states of spinor-1 condensates.
\end{abstract}


\maketitle
\section{Introduction}
\label{sec:introduction}
The availability of atomic Bose-Einstein condensates (BEC) with
spin degrees of freedom have stimulated much recent interest in
their applications to quantum information physics. Atomic spinor-1
condensates (hyperfine spin $f=1$ for each atom) were first realized
by transferring spin polarized BEC prepared in a magnetic trap into
a far-off resonant optical trap \cite{kurn}. More recently, the all
optical route to BEC has opened the door for their wide
accessibility \cite{mike}.
In a simple treatment, a spinor-1 condensate can be described by
a three-component order parameter, one for each of the Zeeman
component of the hyperfine manifold. Early theoretical studies
have clarified rotationally invariant descriptions including
elastic s-wave collisions \cite{ho,ohmi,zhang,goldstein,pu}.
Recent investigations reveal that such a system
also possesses complex ground state structures and can
exhibit novel dynamical effects \cite{pu,ueda2,ho2},
such as fragmentation \cite{james}, spin-mixing
and entanglement \cite{ho,pu,duan}.

In this paper, we further explore
quantum correlation properties such as spin squeezing and
entanglement of a spinor-1 condensate \cite{duan,epr}.
For a spin half $(f=1/2)$ atomic system, a rotationally
invariant Hamiltonian is known to not induce spin squeezing
as the total spin is conserved \cite{kitagawa}.
For a spinor-1 condensate, it was found that its Hamiltonian
becomes rotationally invariant if the single (spatial) mode
approximation is made to its order parameters,
i.e. assuming $\psi_{m_f}(\vec r)=\phi(\vec
r)a_{m_f}$ with the same mode function $\phi(\vec r)$ \cite{ho,pu}.
$a_{m_f}$ is the annihilation operator for atoms in
Zeeman state $m_f=\pm,0$. However, as we show in this work,
various nonlinear processes do occur within different
subspaces of the full SU(3) structure of a spinor-1 condensate,
e.g. we find the existence of spin squeezing in
the isospin sub-group \cite{epr}.

The generation and characterization of squeezing and
entanglement of Bose-condensed atoms has recently emerged as
an active research area. Historically, atomic squeezed states
were first considered for a system of two level atoms.
Even in this simple SU(2) case, it was found that some
operational definitions of spin squeezing can become system dependent.
Depending on the context
in which the concept of squeezing is applied, different definitions
arise \cite{wineland}. Squeezing in
atomic variables was first introduced through reduced
fluctuations in the atomic (Pauli) operators of the system \cite{walls}
such that atomic resonance fluorescence
in the far-field zone is squeezed.
In this case, it is useful and convenient to define the
atomic squeezing parameter according to,
\begin{eqnarray}
\xi_h=\Delta J_i/\sqrt{|\langle J_j\rangle/2|},\hskip 16pt i\neq j\in(x,y,z).
\end{eqnarray}
This definition can be essentially read off from the Heisenberg
uncertainty relation $\Delta J_i\Delta J_j\geq|\langle
J_k\rangle/2|$ for the collective angular momentum components
of the two level atomic system. When in such a
squeezed state $\xi_h<1$, the quantum fluctuation of
one collective angular momentum component becomes lower than
the Heisenberg limited value at the cost of increased fluctuation
in the other component. The general family of two level
atomic states satisfying this criterion was found to be
Bloch states, or SU(2) coherent states. These ``squeezed states"
are obtained by simply rotating in space the collective (Dicke) state
$|J,\pm J\rangle$ \cite{arecchi}. Soon it became clear that
neither $\xi_h$ nor the Bloch states are that useful in
other applications of atomic squeezing.
In particular, for Ramsey oscillatory field spectroscopy,
a new squeezing parameter \cite{wineland2},
\begin{eqnarray}
\xi_R=\sqrt{2J}\Delta J_{\perp}/|\langle \vec{ J}\rangle|
\end{eqnarray}
is called for with
$J_{\perp}$ the angular momentum component normal to the
$\langle\vec{J}\rangle$, i.e. in the direction of the
unit vector $\vec{n}$ along which $\Delta(\vec{n}\cdot\vec{J})$
is minimized. The squeezing condition $\xi_R<1$ is not
straight forwardly determined by the Heisenberg
uncertainty relation. Instead, it is defined by requiring
the improvement of signal to
noise ratio in a typical Ramsey spectroscopy. It was later
shown that the same criterion is also applicable for improving
the phase sensitivity of a Mach-Zehnder interferometer \cite{wineland}.
$\xi_R\equiv 1$ for Bloch states or SU(2) coherent states.
An independent refinement of $\xi_h$ was suggested by Kitagawa
and Ueda \cite{kitagawa} to make it independent of
angular momentum coordinate system or specific
measurement schemes. They emphasized that collective spin
squeezing should reflect quantum correlations between
individual atomic spins and defined a squeezing parameter
\begin{eqnarray}
\xi_q=\Delta J_{\perp}/\sqrt{J/2}
\end{eqnarray}
to measure such correlations.
The factor $J/2$ in the denominator represents the
variance of a Bloch state, which comes from simply adding up the
variance of each individual spin (1/2).
When quantum correlations exist among different atomic spins,
the variance of certain component of the collective spin
can become lower than $J/2$. This leads naturally to the criterion
$\xi_q<1$ for spin squeezing. We note that this definition is
directly related to the spectroscopic definition as
$\xi_R=(J/|\langle J_z\rangle|)\xi_q$.
More recently, a particular type of quantum
correlation, namely the multi-particle entanglement,
becomes important for quantum information physics. A more stringent
criterion for atomic squeezing which combines the quantum
correlation definition with the inseparability requirement
of system density matrix is given by S\o rensen
{\it et al.} \cite{sorensen},
\begin{eqnarray}
\xi_e^2=\frac{(2J)\,\Delta(\vec{n}_1\cdot\vec{J})^2}{\langle
\vec{n_2}\cdot\vec{J}\rangle^2+\langle\vec{n}_3\cdot\vec{J}\rangle^2}<1,
\end{eqnarray}
with the ${\vec n_i}$ being mutually orthogonal unit vectors.
$\xi_e$ is in fact identical to $\xi_R$
along the direction $\langle\vec{n}_1\cdot\vec{J}\rangle=0$.
It was proven rigorously that when $\xi_e<1$, the total state
of the N two level atoms becomes inseparable, i.e. entangled
in a general sense. All three definitions above applies to
a two component (two level) atomic system.
Many complications arise when attempt is made to extend spin
squeezing to a spinor-1 SU(3) system.
Under certain restrictive conditions, $\xi_e$ has been used
recently to discuss a two mode entanglement in a spinor-1
condensate \cite{duan}.

A related problem to spin squeezing is its efficient generation
and detection. In accordance with their respective definitions
for SU(2) systems, several physical mechanisms have been proposed
along this direction. Kitegawa and
Ueda considered a model Hamiltonian
$H_{KU}=\hbar\chi J_z^2$ that can be realized via the
Coulomb interaction between electrons in the two arms of
an interferometer \cite{kitagawa}. Barnett and Dupertuis suggested
that spin squeezing can be achieved in a two-atom system described by
$H_{BD}=i\hbar(g^{\ast}J_{1+}J_{2+}-h.c.)$ \cite{barnett}.
Use of a pseudo-spin two component atomic condensate system has
also been suggested \cite{sorensen}. Recently, two different groups
considered atomic (spin) squeezing and entanglement in a spinor-1 condensate,
under the assumption that one of the component is highly populated
such that quantum properties are important only among the
remaining two sparsely populated components \cite{duan,epr}.
Our aim in this study is to remove such a restrictive condition,
and consider the full quantum correlations within a spinor-1
condensate.

Let us consider a general three component system labeled by
$i,j\in \{+,-,0\}$. For spectroscopic and interferometric
applications the observables of interest are the relative
number of particles $N_i-N_j$ (particle partitioning) and the
corresponding phase differences $\phi_i-\phi_j$ with their
measurements limited by noises $\delta
N_{ij}=\langle[\Delta(N_i-N_j)]^2\rangle^{1/2}$ and
$\delta\phi_{ij}=\langle[\Delta(\phi_i-\phi_j)]^2\rangle^{1/2}$.
For a two-component system, the particle partitioning
becomes the collective angular momentum projection
as $N_+-N_-=2J_z$ in the standard Schwinger representation;
the relative phase becomes the corresponding azimuthal phase
$\phi_z\equiv\phi_+-\phi_-$, which is conjugate to $J_z$.
Quantum mechanically they satisfy
$[J_z,\phi_z]=i$. Thus from $\delta J_z\delta \phi_z \geq 1$
and $\delta N_{+-}=2\langle \Delta J_z^2\rangle^{1/2}$,
we find
$\delta\phi_{+-}\approx\langle \Delta J_y^2\rangle^{1/2}/|\langle
J_x\rangle|$ \cite{kitagawa}. Therefore, for
spin squeezed states, one achieves higher angular
resolution and reduced particle partitioning noise.
In a three component system, we can similarly associate the
three number difference $N_i-N_j$ with three subspace
pseudo-spins (each of spin-$1/2$) $\vec{U}$, $\vec{T}$,
and $\vec{V}$ such that
$N_+-N_-=2T_3, N_+-N_0=2V_3, N_--N_0=2U_3$. The phase differences
can then be similarly expressed in terms of components
$U_{x,y}$, $V_{x,y}$, and $T_{x,y}$. When demanding noise
reduction in such a SU(3) system, we need to consider squeezing
in the three spin-1/2 subsystems. One may naively expect
that results from the above discussed SU(2) squeezing
can be applied to each of the three subsystems,
and collectively, one can simply demand that $\xi_{e}<1$
to be satisfied simultaneously. In reality this does not work as
the three spin-1/2 subsystems do not commute with each other.
This is also the fundamental reason that
makes it difficult to generate and detect quantum
correlations in a full SU(3) system. Furthermore,
due to the above non-commuting nature, the three SU(2) sub-spins
cannot be squeezed independently of each other.
Previous discussions of a spinor-1 condensate entanglement
are always limited to just one SU(2) subspace, usually
in the limit $N_0\sim N$, i.e. one mode is highly populated.
Approximately, this limit destroys the underlying
non-commutative algebra among $(U,V,T)$ and simplifies the
problem to that of a usual two-mode SU(2) system.

One of the major results of this paper is that the
effective Hamiltonian of a general spinor-1 condensate can
be decomposed as
\begin{eqnarray}
H=\hbar\chi_{KU} T_3^2+\hbar\chi_{BD}(U_+V_++h.c.),
\end{eqnarray}
which involves both the Kitagawa-Ueda (KU) and Barnett-Dupertuis (BD)
type of spin squeezing simultaneously. In other words, all three
fictitious spins can indeed be found squeezed in a spinor-1 condensate,
as the above two distinct nonlinearities commute with
each other, and therefore squeeze all three SU(2) subspaces
simultaneously.
We find that the BD type interaction dominates
when $N_0$ is large;
while in the opposite limit the KU type squeezing governs.
For intermediate values of $N_0$
it is necessary to consider a generalized spin squeezing for
the three mode spinor-1 system. To achieve this, we provide a
new criterion for the $U$-$V$ two spin squeezing based on
reduced quantum fluctuations imposed
by the BD type nonlinearity. When such a condition is satisfied,
the state of a spinor-1 condensate as a macroscopic-coherent quantum
object becomes useful for three-mode spectroscopic and interferometric
applications. We further show that this condition also corresponds to a
two-mode entanglement in terms of the Holstein-Primakopf bosonic modes,
and it reduces to previous results in the large $N_0$ limit \cite{duan,epr}.
Squeezing in $T$ spin is particularly useful for quantum
information applications based on collective
(Dicke) states $|J,J_z\rangle$. These states are in fact stationary
in a spinor-1 condensate and can be manipulated via
external control fields \cite{pu}. Since $J_z=N_+-N_-=2T_3$,
such $T$-squeezed states ensure well-defined Dicke states.

In our study of spin squeezing in a spinor-1 condensate
as outlined in this paper,
we present a systematic approach by recognizing the
$(U,V,T)$ pseudo-spin subspaces as the Gell-Mann (quark)
realization of the SU(3) algebra \cite{gellmann}.
Similar recognitions are found useful
in the recent discussions of quantum and semi-classical dynamics
of three coupled atomic condensates \cite{nemoto}, where the
BD-type two-spin squeezing nonlinearity was absent.
Earlier investigations of
three level atomic systems also made efficient use of the
density matrix and expressed atomic Hamiltonians in terms
SU(3) generators \cite{hioe,cook}. The main
difference between our approach (on spinor-1 BEC) and those
earlier studies are the enveloping Weyl-Heisenberg algebra
of the bosonic operators, which leads to subsequently much
larger Hilbert space of the system.
In addition to spin squeezing, we also investigate other
quantum correlation effects, e.g. condensate fragmentation
with the new theoretical framework. We show that
previous theories based upon the SO(2)
rotational symmetry group cannot give a
decomposition of angular momentum operator with nonlinearities
that could easily be considered for spin squeezing. For instance,
neither $H\propto L^2-2N$ \cite{pu} nor $H\propto N(N-1)-A^\dag A$ \cite{ohmi}
can lead to any simple recognition of the nonlinear coupling among
various spin-components.
\section{SU(3) Formulation for spinor-1 BEC}
\label{sec:system}

Under the single-mode approximation \cite{pu}, a spinor-1
condensate is described by the Hamiltonian
\begin{eqnarray}
H=\mu
N-\lambda^{\prime}_sN(N+1)+\lambda^{\prime}_a(L^2-2N),
\label{ham}
\end{eqnarray}
where $N=n_++n_-+n_0$ is the total number of atoms and the
collective angular momentum ($L$) with the familiar raising
(lowering) operator
$L_+=\sqrt{2}(a_+^{\dagger}a_0+a_0^{\dagger}a_-)$
($L_-=L_+^{\dagger}$), $L_z=n_+-n_-$, and $n_i=a_i^{\dagger}a_i$.
$\mu$ is the chemical potential. $\lambda^{\prime}_{s,a}$ are
renormalized interaction coefficients, they are related to various
s-wave scattering lengths and $\phi(\vec r)$ \cite{ho,pu}.
The validity of the single-mode approximation is now reasonably
well understood \cite{pu,duan,yi}, and for ferromagnetic interactions,
it is in fact exact as shown recently \cite{yi}.
$a_{\pm,0}$ can form a similar Schwinger
representation of SU(3) in the following manner,
\begin{eqnarray}
T_+&=&a_+^{\dagger}a_-,\quad T_3=\frac{1}{2}(n_+-n_-),\\
\left(\begin{array}{c} V_+\\U_+ \end{array}\right)&=&a_\pm^{\dagger}a_0,
\quad \left(\begin{array}{c} V_3\\U_3 \end{array}\right)=\frac{1}{2}(n_\pm-n_0),\\
N&=&n_++n_-+n_0, \\
Y&=&\frac{1}{3}(n_++n_--2n_0).
\end{eqnarray}
The linear combinations $X_{\pm}\pm X_{\mp}$ ($X=T,U,V$) together
with $T_3$ and $Y$ resemble the set of eight generators of SU(3)
in spherical representation. $T_{\pm,3}$,$U_{\pm,3}$ and
$V_{\pm,3}$ fulfill commutation relations $[X_+,X_-]=2X_3$ and
$[X_3,X_{\pm}]=\pm X_{\pm}$ of the SU(2) algebra. We call
T-operators the isospin and $Y$ operators the hypercharge only
because of their formal resemblance \cite{gellmann}. $U$ and $V$
subalgebras will be called $U$- and $V$-spin, respectively. We
then have $L_+=\sqrt{2}(V_++U_-)$, $L_z=2T_3$, and
\begin{eqnarray}
L^2 &&=4T_3^2+\frac{1}{2}(N-\epsilon_+)(N-\epsilon_-)\nonumber\\
 &&-2(Y-Y_0)^2+G_Y,\label{l7}\\
G_Y &&= 2(V_+U_++h.c.),
\end{eqnarray}
with $\epsilon_{\pm}=-3/2\pm\sqrt{2}$, and
$Y_0=-N/6-1/4$. For $N\gg 1$ this gives $Y_0\approx -N/6$, which
corresponds to $n_0=N/2$. Using $[T_3,V_{\pm}]=\pm V_{\pm}/2$ and
$[T_3,U_{\pm}]=\mp U_{\pm}/2$ we find $[T_3,G_Y]=0$ consistent
with $[H,L_z]=0$. Hence the Hamiltonian (\ref{ham}) separates
into three commuting parts $H=H_N[N]+H_T[T_3]+H_Y[Y]$.
To our knowledge, this decomposition have not been discussed
before. In Ref. \cite{nemoto}, a model of $H=\chi(T_3^2+3Y^2)$ has been
considered for both the quantum and semi-classical dynamics
of $Y$ as well as for SU(3) coherent states.
We note that the decomposition (\ref{l7}) differs from the
Casimir relation for the two mode case \cite{kris}.
In fact, with the spin
singlet pair operator $A=(a_0^2-2a_+a_-)/\sqrt{3}$ as defined by
Koashi and Ueda \cite{ueda2}, we find $L^2=N(N+1)-A^\dag A$.
\begin{figure}
\includegraphics[width=2.5in]{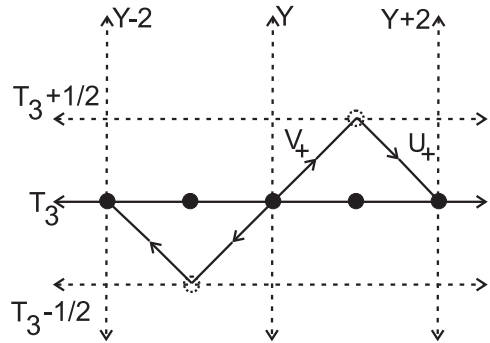}
\caption{The
action of $G_Y$ in $T_3-Y$ space. Any point is coupled only to its
next nearest neighbors along $T_3$-axis through a two step process
$V_+U_+$ on the $Y$-line with $T_3$ unchanged in the end. Note
that $V_+$ and $U_+$ commutes with each other and the conjugate
process is also shown.} \label{fig1}
\end{figure}

Denote the simultaneous eigen-states
of commuting operators ($N,Y,T_3$) as $|N,T_3,Y\rangle$,
we find
\begin{eqnarray}
&&\left(\begin{array}{c} V_+\\U_+ \end{array}\right)|N,T_3,Y\rangle\nonumber\\
&&=\sqrt{\left(\frac{N}{3}-Y\right)\left(\frac{N}{3}\pm
T_3+\frac{Y}{2}+1\right)}|N,T_3\pm\frac{1}{2},Y+1\rangle,\nonumber
\end{eqnarray}
i.e., $G_Y$ only couples next nearest neighbors along the $Y$-axis
through off-axial hopping as depicted in Fig. \ref{fig1}.
Perhaps it is not surprising that operators $T_{\pm}$, $U_{\pm}$,
and $V_{\pm}$ are simply the off-diagonal elements of the
single particle density operator $\rho_{\mu\nu}=a_{\mu}^{\dagger}a_{\nu}$,
while $N$, $T_3$, and $Y$ are related to the diagonal elements.
$T_{\pm},V_{\mp},U_{\pm}$ all raise and lower the $T_3$ value
by ($1$ or $1/2$).

\begin{figure}
\includegraphics[width=3.in]{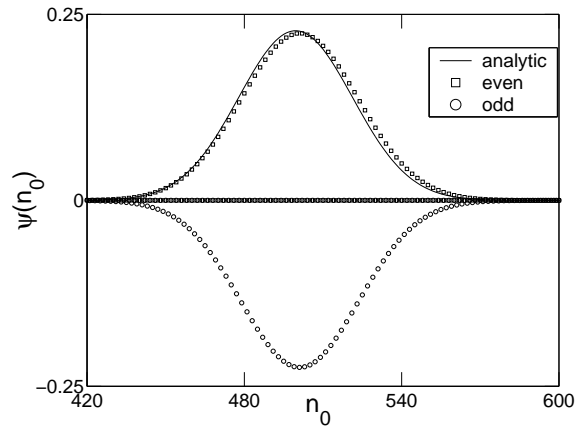}
\caption{The ground state expansion coefficients
$\psi(Y)\equiv\psi(n_0)$ as a function of $n_0$ for $N=1000$ atoms
in the $T_3=0$ block. The solid curve is the approximate analytical
result Eq. (15) without
any fitting parameters while the other curves are obtained by
an exact diagonalization procedure.} \label{fig2}
\end{figure}

As an example we consider the simple case of the $T_3=0$ block
along the line in Fig. \ref{fig1} of the
Hamiltonian (\ref{ham}) for $\lambda^{\prime}_a<0$,
the ferromagnetic case (as in $^{87}$Rb \cite{mike}).
The polar case of $\lambda^{\prime}_a>0$ (as in $^{23}$Na BEC \cite{kurn})
has been discussed in Ref. \cite{ho}. Dropping the constant $H_N[N]$
and write in units of $|\lambda^{\prime}_a|$, the Hamiltonian becomes
\begin{eqnarray}
H=2(Y-Y_0)^2-2t_Y \left(|Y+2\rangle\!\langle Y|+h.c.\right),
\end{eqnarray}
where $t_Y=(N/3-Y)(N/3+Y/2+1)$. Following the tight-binding
procedure for the restricted two-mode case ($+$ and $-$)
discussed in Ref.
\cite{ho2}, its eigenstates can be found by determining the
$\psi(Y)$ of $|\psi\rangle=\sum_Y\psi(Y)|Y\rangle$ through a
difference equation
\begin{eqnarray}
E\psi(Y)&=&2(Y-Y_0)^2\psi(Y)\nonumber\\
&&-2[t_{Y-2}\psi(Y-2)+t_Y\psi(Y+2)].
\end{eqnarray}
In the continuum limit and up to the first order in $O(|Y|/N)$,
the equivalent differential form becomes
\begin{eqnarray}
\left(\frac{E}{8Y_0^2}+1\right)\psi=-2\frac{\partial^2\psi}{\partial
Y^2} -\frac{1}{Y_0}\frac{\partial \psi}{\partial Y}+
\frac{(Y-Y_0)^2}{4Y_0^2}\psi.
\end{eqnarray}
Its ground state is therefore
\begin{eqnarray}\label{hyper_grnd}
|\psi\rangle=\sum_Y {
\exp{\left({-\frac{\sqrt{2}(Y-Y_0)^2}{8|Y_0|}+\frac{Y-Y_0}{4|Y_0|}}\right)}
\over (\pi|Y_0|/\sqrt{2})^{1/4} } |Y\rangle,
\end{eqnarray}
which gives a diagonal $\langle\rho_{\mu\nu}\rangle$ with $\langle
n_0\rangle=N/2$ and $\langle n_{\pm}\rangle=N/4$, i.e. a
fragmented state \cite{ho2,james}. To check the validity of this
approximate analytical result, we also solved the same problem
for $N=10^3$ within the $T_3=0$ block by an exact diagonalization
procedure. The results are compared in Fig. \ref{fig2}.
We see that without any fitting parameter the
analytical result agrees well with the exact numerical result.
Since the Hamiltonian is block diagonal in even $n_0$ and odd
$n_0$ spaces, we find two degenerate ground states with even and
odd $n_0$ components respectively. These even-odd ground states
display opposite phases and can form a Schrodinger cat state \cite{frag}.
The approximate result here applies for a value of even $N$,
which leads to an even $n_0$ within the $T_3=0$ block.
Hence, only the even $n_0$ block of the
Hamiltonian is considered. We will also show below that
under the single-mode approximation, the exact ground state
for $N\gg 1$ is generally a fragmented state with $\langle n_0\rangle=N/2$.
More detailed studies and implications of the cat-like ground
state in the even-odd number blocks and with their respective
phases will be explored further and results presented elsewhere.
We note that such cat states separated in the angular
momentum $L_z$ have been found in studies of Josephson
type coupled condensates \cite{cat1,cat2}. In a spinor-1
condensate as considered here, we find that the
dynamical behavior can be characterized by $Y=N/3-n_0$, which can be
expressed as $Y=2(U_3+V_3)/3$. Since the azimuthal phases are
conjugate to angular momentum z-projection operators,
the cat-like ground states predicted here resemble
the angular momentum cat states of a two-component condensate in
its conjugate phase spaces. Finally we note that the symmetry
point $Y_0$ can be adjusted by external control fields
which contribute terms proportional to
$n_{\pm,0}$ to the Hamiltonian, and can be absorbed into the
$(Y-Y_0)^2$ term through a new $Y_0$ as
$n_{\pm}=N/3+Y/2\pm T_3$ and $n_0=N/3-Y$.

We have now seen that the Hamiltonian of the system describes an
effectively one dimensional dynamics along the $Y$-axis, similar
to that of a diffusive random walk process but now with an
attractor (for $\lambda_a^{\prime}<0$) $Y_0$. Hence, we expect
$Y_0$ influence population dynamics in a similar manner it effects
the fragmentation. For $T_3=0$, it is known that populations
oscillate around time-averaged values $n_0=N/2$ and $n_{\pm}=N/4$
which are the same as the results we found for fragmented ground
states.  We conclude that steady-state values of population
oscillations as well as fragmentation is determined by the
hypercharge symmetry point $Y_0$, which can be shifted by external
fields.
\section{Two-spin and isospin squeezing}
The form of $G_Y$ suggests the existence of two mode squeezing as
was also noted recently by Duan {\it et al.} \cite{duan}, who
studied a spinor-1 condensate initially prepared in the Fock state
with only $m_f=0$ state populated. During the time
when the total number of excitations into states $m_f=\pm 1$ are negligible,
the spin mixing term ($G_Y$) in the Hamiltonian simply reduces
to a two-mode squeezing nonlinearity
via $\langle n_0\rangle(a_+^{\dagger}a_-^{\dagger}+h.c.)$.
This creates a continuous variable type entanglement, or
mode-entanglement in the second quantization form.
In order to relate it to measurable spectroscopic spin squeezing
and particle entanglement, Ref. \cite{duan} first showed that
in the low excitation limit, the two-mode entanglement criterion
can also be expressed in terms
of spin squeezing parameters for $L_{x,y}$. In order to use
the two-level SU(2) definition for spin squeezing of $L_{x,y}$,
new pseudo-spins $J_{\pm}$ was introduced within the
the two level subsystems $|+1\rangle\pm|-1\rangle$ and $|0\rangle$.
They found that when
$L_z\approx 0$, the system Hamiltonian becomes effectively
$H=\lambda_a^{\prime}(L^2-2N)\approx\lambda_a^{\prime}
(L_x^2+L_y^2)\sim (J_{+x}^2+J_{-y}^2)$, which causes
each spin 1/2 subsystem to be
squeezed via the single axis twisting scheme.
For the independent single-axis twisting scheme to work
efficiently in achieving substantial spin squeezing,
the commutator $[J_{+x/y},J_{-x/y}]=(T_--T_+)/4\propto T_y$,
needs to be small. Hence, squeezing in the isospin is essential
to achieve this two-mode squeezing goal.
Without it, large quantum fluctuations in $T_y$ would
destroy the two mode squeezing. Unfortunately, both the relationship
between the two-mode squeezing and spin squeezing as well as
the interpretation in
terms of a dual single axis twisting fails to be adequate for
higher excitations under more realistic situations. Indeed, for
the extreme opposite case of $n_0\ll n_{\pm}\sim N/2$, the
Hamiltonian describes a single-mode amplitude squeezing as
it reduces to $G_Y\sim (a_0^{\dagger 2}+h.c.)$. Anywhere in between
of these two extreme limits, we propose a new type of squeezing,
the two-spin squeezing as a generalization of single spin squeezing by
taking into account quantum correlations for mode-entanglement
applications. We first note that the two extreme types of squeezing
in $G_Y$ can be handled at arbitrary levels of excitation by
introducing a new two-spin squeezing operator via
\begin{eqnarray}
K_+&=&V_+U_+\sim\left\{\begin{array}{cc}
a^{\dagger}_+a^{\dagger}_-,&n_0\gg n_{\pm},\\
a_0^2,&n_0\ll n_{\pm},\end{array}\right .\\
K_-&=&K_+^{\dagger},
\end{eqnarray}
with $[K_-,K_+]=2K_3$. The squeezing mechanism in
Hamiltonian (\ref{ham}) is now understood to be a generalized
Barnett-Dupertuis (BD) type squeezing via the $V_+U_++h.c.$
nonlinearity in $G_Y$. This is significantly more complicated than
the two bosonic mode squeezing as the two spins $U$ and $V$
have a non-commuting algebra. The mode-entanglement of
approximate bosonic modes $a_\pm^{\prime}=a_\pm
a_0^{\dagger}/\sqrt{\langle n_0\rangle}$ of Ref. \cite{duan}
can in fact be generalized to mode-entanglement between exactly bosonic
Holstein-Primakoff modes \cite{holstein} $a_{x=u,v}$ defined
through $X_+=a_x^{\dagger}\sqrt{S_x-N_x}$ and $X_3=N_x-S_x/2$,
in the spin $S_x/2$ realization of corresponding SU(2) algebras of
$U$ and $V$ spins with $N_{x}=a_{x}^{\dagger}a_{x}$. The
squeezing treatment with the exact bosonic modes
$a_U$ and $a_V$ remains to be more complicated than
the usual two bosonic mode squeezing as it also suffers
from the underline non-commutating algebra.
This representation
reduces to the usual SU(1,1) two-mode squeezing or amplitude
squeezing in the appropriate $n_0$ limits. At low excitations when
$n_0\approx N$, we have $X_3\approx -N/2$, $S_{x}\approx\langle
n_0\rangle$, and $N_{x}\approx 0$. In this case, $X_-\approx
\sqrt{\langle n_0\rangle}a_{x}$ and $G_Y=2\langle n_0
\rangle(a_v^{\dagger}a_u^{\dagger}+h.c.)$ demonstrates the
two-mode [SU(1,1)] squeezing as in Ref. \cite{duan}. In the
large $n_0$ scheme of Ref. \cite{duan}, such modes are sparsely
populated since, $a_+^{\prime\dagger}a_+^{\prime}=n_+(1+n_0)/n_0$.
In the opposite case of large $n_{\pm}$, we are in the strong
excitation regime with $N_v\approx 1, S_{v/u}\approx \langle
n_{\pm}\rangle$ which gives effective modes to be
$a_{\pm}^{\dagger}a_0/\sqrt{n_{\pm}}$ with large occupations.

In order to define two-spin squeezing introduced via the
$K$-operators in a similar way to the two-mode bosonic squeezing,
we introduce Hermitian quadrature operators
\begin{eqnarray}
X_{u}^{\alpha}&=&(e^{i\alpha}U_-+h.c.)/\sqrt{2}, \\
X_v^{\alpha}&=&(e^{i\alpha}V_-+h.c.)/\sqrt{2},\\
Q_{+}^{\alpha}&=&(X_v^{\alpha}+X_{u}^{\alpha})/2,\\
Q_-^{\alpha+\pi/2}&=&(X_v^{\alpha+\pi/2}-X_u^{\alpha+\pi/2})/2.
\end{eqnarray}
From $\vec{J}_{\pm}=\vec{V}-\vec{U}$ and
$J_{\pm 3}=(3Y/2\pm T_x)/2$ we find
$Q_{\pm}^{\alpha}=\vec{n}(\alpha)\cdot\vec{J}_{\pm}$ with
$\vec{n}(\alpha)=(\cos{\alpha},\sin{\alpha},0)$. If $U$ and $V$
were uncorrelated, their respective quantum noises would
contribute to that of $J_{\pm}$ additively.
Existence of quantum correlations between the $U$- and $V$-spins
would reduce the quantum fluctuation in $J_{\pm}$. Thus,
the ($J_{\pm}$) spin squeezing is achieved by two-spin ($U$-$V$)
squeezing. From $[U_-,V_-]=0$, we find that
\begin{eqnarray}
&&(\Delta Q_+^{\alpha})^2+(\Delta Q_-^{\alpha+\pi/2})^2\nonumber\\
&&=\sum_{a=u,v}[(\Delta X_a^{\alpha})^2+(\Delta
X_a^{\alpha+\pi/2})^2]
+{\cal C}_{uv}^{\alpha},
\end{eqnarray}
with the $U$-$V$ correlation function
\begin{eqnarray}
{\cal C}_{uv}^{\alpha}=e^{-2i\alpha}\langle V_+,U_+\rangle/2+c.c.
\end{eqnarray}
denote the correlations among $U$-$V$ spins to the quadrature noise,
which reduces the uncertainty bound when two spin squeezing occurs.
We find a lower bound for the quadrature noise
$\sum_{a=u,v}[(\Delta X_a^{\alpha})^2+(\Delta
X_a^{\alpha+\pi/2})^2]$, by noting that
$[X_v^{\alpha},X_v^{\alpha+\pi/2}]=2iV_3$,
$[X_u^{\alpha},X_u^{\alpha+\pi/2}]=-2iU_3$,
and $|\langle
U_3\rangle|+|\langle V_3\rangle |\ge |\langle U_3+V_3\rangle
|=3|\langle Y\rangle|/2$. We finally find
\begin{eqnarray}
(\Delta Q_+^{\alpha})^2+(\Delta Q_-^{\alpha+\pi/2})^2\ge 3|\langle
Y\rangle|/4+{\cal C}_{uv}^{\alpha}.
\end{eqnarray}
Therefore, taking into
consideration the important spin-spin correlation between
different particles similar to the $1/2$ case \cite{kitagawa},
we can introduce the $U$-$V$ squeezing condition as
\begin{eqnarray}
\xi_{uv}^{\alpha}=\frac{\Delta (Q_+^{\alpha})^2+(\Delta
Q_-^{\alpha+\pi/2})^2}{|\langle Y\rangle|}< 3/4,
\label{gs}
\end{eqnarray}
similar to the continuous variable system \cite{con}.
This is the major result of our paper on the two [SU(2)]
spin squeezing within the SU(3) of a spinor-1 condensate.
The significance of spin-spin correlation function to spin
squeezing and entanglement for a two-mode system was
previously discussed in Ref. \cite{agarwal}, where they
showed that a negative, finite correlation parameter causes
spin squeezing and entanglement of the atomic states.
With the Holstein-Primakoff relations, it
is straight forward to show this condition contracts into
$(\xi_+^{\alpha})^2+(\xi_-^{\alpha+\pi/2})^2<2$ when $n_0\rightarrow N$.
Thus Eq. (\ref{gs}) generalizes the two-mode entanglement criterion
$(\xi_+^{\alpha})^2+(\xi_{-}^{\alpha+\pi/2})^2<2$ at low excitations
\cite{duan} to arbitrary levels of excitation
for two spin squeezing. For completeness,
we note the squeezing parameter for $J_{\pm}$-spins are
\cite{duan}
\begin{eqnarray}
(\xi_{\pm}^{\alpha})^2={N\langle (\Delta Q_{\pm}^{\alpha})^2 \rangle
\over \langle Q_{\pm}^{\alpha+\pi/2}\rangle^2+\langle J_{\pm
3}\rangle^2}, \label{bs}
\end{eqnarray}
while the Heisenberg uncertainty relation gives $\Delta
Q_{\pm}^{\alpha}\Delta Q_{\pm}^{\alpha+\pi/2}\geq|\langle J_{\pm
3}\rangle|/2$. For many particle entanglement of three level
atoms, the criterion is given by either $\xi_{\pm}<1$.

The $U$-$V$ squeezing discussed above displays existence of
nonlinear interactions within/among $T$, $U$, and $V$ subspace of
(\ref{ham}). One may also contemplate for a one-axis isospin
twisting (through $T_3^2$) of the particular form of $L^2$ (\ref{l7}).
However, the dynamics of spinor-1 BEC becomes
considerably more complicated because of off-axis hopping processes along
the hypercharge axis (as in Fig. 1). Due to the non-commutativity of
sub-spin systems ($U,V,T$), squeezing and entanglement appears
even without essentially any axis-twisting. In fact, even when
$T_3=0$, squeezing within the isospin subgroup can still happen
as the $U$-$V$ two-spin squeezing interaction would redistributes
the noise also for the isospin subspace, in addition to the $U$-$V$ spin
space. To appreciate this fact, let us consider the rotation operator
involving only $U$-$V$ spins and employ the SU(2) disentangling
theorem to obtain
\begin{eqnarray}
R[\zeta]=e^{\zeta L_+-\zeta^{\ast}L_-}=e^{\eta
L_+}(1+|\eta|^2)^{L_z}e^{-\eta^{\ast}L_-},
\end{eqnarray}
with $\eta=\zeta\tan{\zeta}/|\zeta|$. Using
$[V_+,U_-]=T_+$ and $[V_+,T_+]=[U_-,T_+]=0$, we find
\begin{eqnarray}
e^{\eta L_+}=e^{\eta\sqrt{2}V_+}e^{\eta\sqrt{2}U_-}e^{-\eta
T_+/\sqrt{2}}.
\end{eqnarray}
Hence, we arrive at
\begin{eqnarray}
R[\zeta]=e^{\sqrt{2}\eta V_+}e^{\sqrt{2}\eta
U_-}R_T[\eta]e^{-\sqrt{2}\eta^{\ast}U_+}e^{-\sqrt{2}\eta^{\ast}V_-},
\end{eqnarray}
with a rotation operator within isospin space via $R_T=\exp{(-\eta
T_+/\sqrt{2})}(1+|\eta|^{2})^{2T_3}\exp{(-\eta^{\ast}T_-/\sqrt{2})}$.
This result reflects the nature of Euler-angle rotations in three
dimensions for a spin-1 system. We thus conclude that squeezing
in $\vec{J_{\pm}}$ through redistributing the noise via rotations
is always accompanied by a redistribution of the noise in the
isospin subspace. Squeezing and many particle entanglement via
the isospin can be checked using the usual spin squeezing
criterion, which for both
$T$-squeezing and the above derived $U$-$V$
squeezing are independent of their respective initial conditions.
Hence, we have now greater freedom to consider a suitably
prepared spinor-1 condensate to achieve many-particle and/or mode
entanglement for quantum information applications as well as
various type spin squeezing for atom interferometry and spectroscopy
applications in the long time limit with more macroscopic
populations in all $f=1$ three component states can occur.
In the limiting case discussed before
either $n_0\sim N$ or $n_{\pm}\sim N$ is required to be large,
the quantum states (modes) of interest are always sparsely populated.
More generally, one can use Raman coupled laser pulses on
a spinor-1 condensate to generate states with arbitrary
populations in each mode and with arbitrary initial phases.
This allows then for the consideration of
stationary states in the fully quantum mechanical framework
for their use in squeezing-entanglement applications.

\section{Results and Discussions}
We now present some results on the
numerical investigation of isospin squeezing. If
the condensate atom number $N$ is fixed, a generic state
$|\psi(0)\rangle=(\alpha_0a_0^{\dagger}+\alpha_-a_-^{\dagger}
+\alpha_+a_+^{\dagger})^N|0,0,0\rangle/\sqrt{N!}$,
can be prepared with Raman pulses \cite{kurn},
where $|0,0,0\rangle$ is the vacuum in the Fock basis
$|n_0,n_-,n_+\rangle$ and $\alpha_j=|\alpha_j|e^{i\delta_j}$
complex. Using $m=n_+-n_-$, we can write
$$|\psi(0)\rangle =
\sum_{mk}\psi_{Nmk}(\vec{\alpha})\left.\left|2k,\frac{N-m}{2}-k,
\frac{N+m}{2}-k\right\rangle\right.,$$
where $\vec{\alpha}=(\alpha_0,\alpha_-,\alpha_+)$,
$k=0,1,\cdots,(N-|m|)/2$ for even $N+m$, $2k=1,3,\cdots,(N-|m|)$
for odd $N+m$, and
$$\psi_{Nmk}=\sqrt{C_N^{2k} C_{N-2k}^{{N-m\over 2}-k} }
\alpha_0^{2k}\alpha_-^{{N-m\over 2}-k}\alpha_+^{{N+m\over 2}-k},$$
where $C_n^m=\left(\begin{array}{c} n\\m \end{array}\right)$
denotes the binomial coefficient.
The basis transformation coefficients
between angular momentum and Fock sates are available
from Ref. \cite{wu}, written in more compact forms as
\begin{eqnarray}
|lm\rangle &=& \sum_k G_{lmk}\left.\left|2k,\frac{N-m}{2}-k,
\frac{N+m}{2}-k\right.\right\rangle,
\end{eqnarray}
with
$$G_{lmk}=2^ks_l\sum_{r}\frac{(-1)^r}{4^r}\left[\begin{array}{c}
Nlm\\kr\end{array}\right]$$
and the symbolic notation
\begin{eqnarray}
\left[\begin{array}{c} Nlm \\kr\end{array}\right] &&=
\sqrt{C_{2r}^r C_{2k}^{2r} C_l^{2k-2r}
C_{N-2k}^{l-2k+2r} C_{N-l-2r}^{(N-l)/2-r}}\nonumber\\
&&\times C_{l-2k+2r}^{(l-m)/2-k+r}/
\sqrt{C_{N-2k}^{(N-m)/2-k} C_{2l}^{l-m} }.
\end{eqnarray}
We note $l=N,N-2,\cdots,N-2[N/2]$
with $[n]=n,n-1/2$ for $n=$ even, odd,
and $r=\max[0,k-(l-|m|)/2],\cdots,\min[k,(N-l)/2]$,
$m=0,\pm1,\cdots,\pm l$. $k=0,1,\cdots,(N-|m|)/2$ for $l+m=$ even
and $2k=1,3,\cdots,(N-|m|)$ for $l+m=$ odd. The normalization
is given by
\begin{eqnarray}
{1\over s_l^2}=\sum_{j=0}^{(N-l)/2}\frac{1}{4^j}
C_{2j}^j C_{(N+l)/2-j}^l.
\end{eqnarray}
Simpler analytic results exist for special cases, e.g.
\begin{eqnarray}
G_{Nmk}=2^k \sqrt{C_N^{2k} C_{N-2k}^{(N-m)/2-k} }/
\sqrt{C_{2N}^{N-m}},
\end{eqnarray}
which takes an asymptotic form
$ G_{N0k}\simeq \sqrt{ C_N^{2k} /2^{N-1}}$
when $N\gg 1$. We also find
\begin{eqnarray}
\langle n_0\rangle = \frac{1}{2^{N-1}}\sum_{k=0}^{N/2}
C_N^{2k} (2k)=N/2,
\end{eqnarray}
the same result as obtained previously from
(\ref{hyper_grnd}).

\begin{figure}
\includegraphics[width=3.25in]{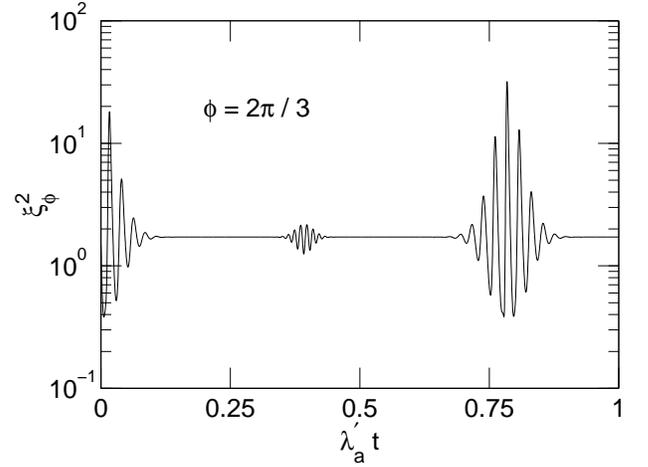}
\caption{Time-dependent squeezing parameter at $\phi=2\pi/3$ for
$N=100$ atoms, $P_0=1/3$, and $\theta=\pi/2$.} \label{fig3}
\end{figure}

\begin{figure}
\includegraphics[width=3.25in]{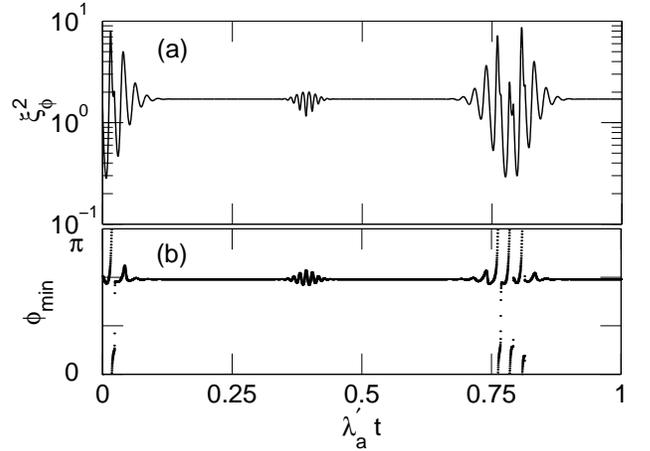}
\caption{(a) The same as in Fig. 2 but for the optimized squeezing
parameter; (b) the optimal angle $\phi_{\rm min}$ which maximizes
squeezing as in (a). } \label{fig4}
\end{figure}

These expressions allow us to express the initial state as
$|\psi(0)\rangle=\sum_{lm}\psi_{lm}(0)|lm\rangle$,
with
\begin{eqnarray}
\psi_{lm}(0)&=&\Lambda_{Nm}(\alpha_-,\alpha_+)\sum_k\eta^k(\vec{\alpha})
G_{Nmk}G_{lmk}, \\
\Lambda_{Nm}&=&\sqrt{C_{2N}^{N-m} }\alpha_-^{(N-m)/2}\alpha_+^{{N+m}/2},
\end{eqnarray}
and $\eta={\alpha_0^2}/({2\alpha_-\alpha_+})$. When $\eta=1$, it
becomes an eigenstate of Hamiltonian (\ref{ham}) as $\sum_k
G_{Nmk}G_{lmk}\to \delta_{Nl}$. This generalizes the stationary
state of \cite{pu} into the quantum regime. The condition for
stationarity becomes $2\delta_0=\delta_-+\delta_+$ and
$|\alpha_-|+|\alpha_+|=1$ (since
$|\alpha_0|^2+|\alpha_-|^2+|\alpha_+|^2\equiv 1$).
For the special case of $\delta_j=0$ and $\alpha_-=\alpha_+$
we obtain $\alpha_0^2=1/2$, in complete agreement with earlier
results \cite{pu}. By defining $P_j=|\alpha_j|^2$ as spin component
populations, we find that stationary states require $P_0=1/2$
whenever $P_-=P_+$. This is, however, not sufficient without
establishing the phase constraint found above, which becomes
particularly useful as it provides for more freedom in state
preparation using Raman coupled laser fields. As an example, we
now consider isospin squeezing with the same form of initial
states as in Ref. \cite{pu} for
$\alpha_0=\sqrt{P_0}\,e^{i\theta/2}$ and
$\alpha_{\pm}=\sqrt{1-P_0}$. This gives $\langle
T_x(0)\rangle=N(1-P_0)/2$ as the only non-vanishing isospin
component (at $t=0$). The population in the $m_f=0$ component
then acts as a knob between the two extreme squeezing type
discussed earlier as well as between the $G_Y$
and $T_3$ terms. In the special case of within the
$T_3=0$ block, we find that the
dynamics of the system is determined only along the hypercharge
$Y$ axis. Previous study in Ref. \cite{duan} with initial
state $|0,N_0,0\rangle$ results in spin-mixing dynamics,
due to which $N_0$ was found to quickly reduce to some
value without further oscillations or recovery.
In our scheme, we find $n_{0,\pm}$ all exhibits collapse
and revival patterns, so does $Y$ as $Y=N/3-n_0$.
Even for the $T_3=0$ block, we have seen redistribution
of noise among the $U$-$V$ components affects
fluctuations in the isospin as well. The squeezing parameter
\begin{eqnarray}
\xi_{\phi}^2={{N\langle\Delta(T_y^{\prime})^2\rangle}\over {\langle
T_x^{\prime}\rangle^2+\langle T_z'\rangle^2} },
\end{eqnarray}
 is analogous to
$\xi_p^\alpha$ (\ref{bs}) but for isospin $T^{\prime}=R[\phi]T$
after rotated around $x$-axis by an angle $\phi$. Isospin
squeezing is then characterized by $\xi_\phi<1$. At $\phi=2\pi/3$,
this occurs after a very short time (see Fig. \ref{fig3}). It is
especially interesting to note that $\xi_\phi$ exhibits collapse
and revival patterns. The optimal angle $\phi_{\rm min}$ for
maximal squeezing (minimal $\xi_\phi$) \cite{kris,sorensen,mol} is
shown in Fig. \ref{fig4}. It oscillates around its time-averaged
value $\approx 2\pi/3$. In general, we find $\xi_\phi$ achieves
its minimum sooner and the minimum is smaller with decreasing
values of $\theta$ or increasing values of $P_0$.

This effect is clearly unique to three-mode systems.
In usual population spectroscopy (e.g. Ramsey type) or in
interferometry (e.g. of Mach-Zehnder type) for a two-mode system,
particle partitioning noise and phase sensitivity can only be controlled by
the modes involved directly. Here, the $m_f=0$ mode actually does not belong
to the isospin group, yet it still influences the isospin noise properties.
In contrast to the two-mode result $N_{\pm}=J\pm J_z=N/2\pm 2T_3$,
a three-mode system has $N_{\pm}=N/3+(U_3+V_3\pm 2T_3)/2$.
A direct measurement of $N_{+}$ or $N_{-}$ will uncover all
noise terms due to quantum correlations among the
various spin components.
A measurement of $N_+-N_-$, on the other hand,
is similar to the two-mode case
as the result is only affected by the noise in the isospin.
When $T_3=0$, the influence of the $m_f=0$ mode population
is reflected in the two-spin squeezing interaction
between the $U$- and $V$-spins, which in turn also
redistributes the noise in isospin.

\begin{figure}
\includegraphics[width=3.25in]{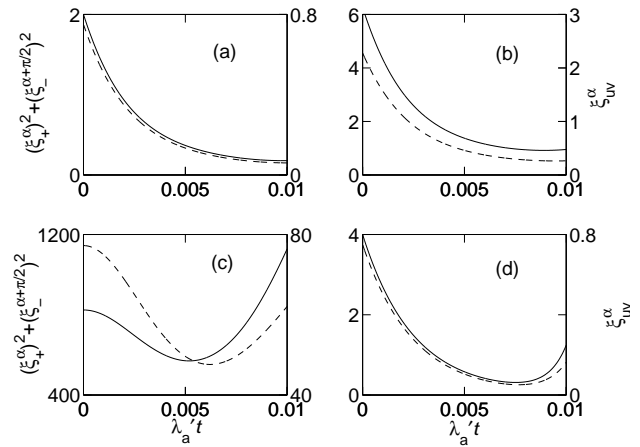}
\caption{Time-dependent $U$-$V$ squeezing parameter (dashed curve)
and two-mode entanglement criterion (solid curve) for $N=100$
atoms initially prepared in a Fock state of
$\psi(0)=|N_-,N_0,N_+\rangle$:  $|0,100,0\rangle$ in (a),
$|1,98,1\rangle$ in (b), $|25,50,25\rangle$ in (c), and
$|50,0,50\rangle$ in (d). } \label{fig5}
\end{figure}

\begin{figure}
\includegraphics[width=3.25in]{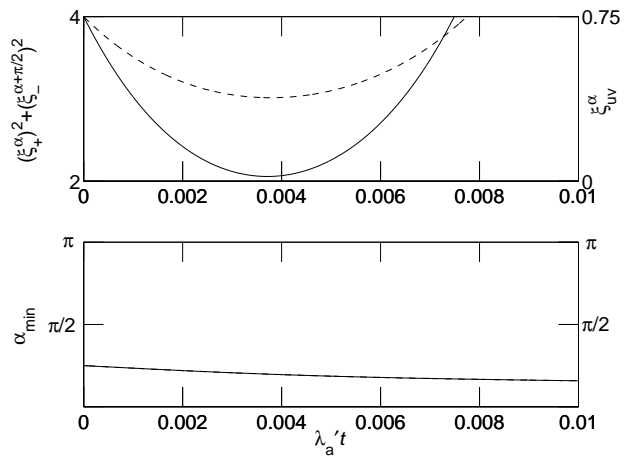}
\caption{Same as Fig. \ref{fig5} but now for
the initial state $\psi(0)=|25,0,75\rangle$.}
\label{fig6}
\end{figure}

In Fig. \ref{fig5}, results of two-spin squeezing are shown for
various initial Fock states $|N_-,N_0,N_+\rangle$ of a spinor-1
condensate. The lack of oscillations in Fig. 1(a) is due to
non-oscillatory behavior of $n_0$ for the particular initial
conditions used here. The solid curves are for the two-mode
entanglement criterion of Ref. \cite{duan}, valid only when
$N_0\gg N_{\pm}$. We see that when the initial states are such
that $N_{\pm}$ modes are not near empty, the achievable two-spin
or two mode squeezing essentially diminishes. However, there is a
also turning point, when squeezing is again recovered if $N_{\pm}$
becomes significantly populated. Hence, we have found a new
squeezing regime when the initial conditions are such that
$N_{\pm}\gg N_0$. The results are almost equivalent to the case
$N_0\gg N_{\pm}$ considered in Ref. \cite{duan}. This new initial
condition generates the two-mode entanglement via two spin
squeezing between the $U$-$V$ spin modes, i.e. between the
holstein-Primakoff bosons. It should be noted that the two-mode
entanglement criterion in terms of spin squeezing parameters
$(\xi_{\pm}^{\alpha})^2$ has been derived for $N_0\gg N_{\pm}$ in
Ref. \cite{duan}. We show here that this criterion is also
satisfied in the opposite case of $N_{0}\ll N_{\pm}$. This
observation emphasizes that the $U$-$V$ squeezing criterion and
the corresponding mode-entanglement can be sought for other
initial conditions when the criterion of Ref. \cite{duan} is no
longer applicable. For that aim, we consider an initial state
$|25,0,75\rangle$ as shown in Fig. \ref{fig6}, where the $U$-$V$
squeezing is indeed found.

\section{Conclusion}
\label{sec:conclusion}
We have provided a comprehensive treatment of
quantum correlations in a spinor-1 condensate. Although no
nonlinear interaction is apparent in the spinor condensate Hamiltonian
when single mode approximation is made, interesting quantum
correlations do develop within subgroups of the SU(3) system.
We have analyzed a spinor-1 condensate in terms of its $T$-, $U$-, and
$V$-spin components. We have found and characterized squeezing within
one particular subgroup, similar to that of the isospin structure
and we have numerically investigated its dynamics in terms of
collapses and revivals. We have developed the
$U$-$V$ spin squeezing as a
generalization of the often adopted spin (1/2) squeezing
\cite{kitagawa} to two-spin squeezing. Its relation to
mode-entanglement \cite{duan} in the Holstein-Primakoff
representation is also pointed out. We have presented new results
for condensate
fragmentation and spin-mixing phenomena in terms of the hypercharge
symmetry and provided general phase-amplitude conditions for
stationary states in the full quantum regime.

In a typical experiment, a small magnetic field gradient may
be available \cite{stenger}, which results in an effective
Hamiltonian \cite{ho2}
$H_B=\alpha(T_++T_-)+\beta T_3^2-\gamma_B T_3$,
instead of (\ref{ham}), with $\alpha$, $\beta$, and $\gamma_B$
various renormalized parameters. In this case isospin squeezing still
occurs through the one-axis twisting nonlinearity \cite{kitagawa}.

Spin squeezing parameters can be measured directly by the
interferometry or Ramsey spectroscopy \cite{kitagawa}.
Alternatively, the isospin ($T$) squeezing in spinor-1
condensate can
also be observed experimentally with light scattering. Using Raman
coupled laser fields, an interaction of the type
$H_R=g(T_+J_-+h.c.)$ can be engineered \cite{wu,wang}, where
$J_-=\sqrt{2}(a_La^{\dagger}_S+a_L^{\dagger}a_A)$ is an angular
momentum operator, with $a_A,a_S,a_L$ the annihilation operators
for anti-Stokes, Stokes, and pump photons. The interaction $H_R$
allows for the mapping of spin correlations into photon
correlations as the total angular momentum $T_3+J_z$ is conserved.
The solutions for $J_-(t)$ depend on the initial conditions
$J_-(0)$ and $T_-(0)$ \cite{wu,wang}. Therefore, the quadrature
operators of scattered
photons are directly related to initial condensate spin
quadratures and a homodyne measurement for Stokes parameters of
the Raman field can reveal isospin squeezing \cite{ozgur}.

\section{Acknowledgements}
We thank Dr. Y. Su for helpful discussions.
This work is supported by the
NSF grant No. PHY-9722410 and by a grant from
the National Security Agency (NSA), Advanced Research and
Development Activity (ARDA), and the Defense
Advanced Research Projects Agency (DARPA) under Army Research Office
(ARO) Contract No. DAAD19-01-1-0667.



\end{document}